# On the cooperativity of association and reference energy scales in thermodynamic perturbation theory

Bennett D. Marshall

*ExxonMobil Research and Engineering, 22777 Springwoods Village Parkway, Spring TX 77389*

**Abstract**

Equations of state for hydrogen bonding fluids are typically described by two energy scales. A short range highly directional hydrogen bonding energy scale, as well as a reference energy scale which accounts for dispersion and orientationally averaged multi-pole attractions. These energy scales are always treated independently. In recent years, extensive first principle quantum mechanics calculations on small water clusters have shown that both hydrogen bond and reference energy scales depend on the number of incident hydrogen bonds of the water molecule. In this work we propose a new methodology to couple the reference energy scale to the degree of hydrogen bonding in the fluid. We demonstrate the utility of the new approach by showing that it gives improved predictions of water-hydrocarbon mutual solubilities.

Bennettd1980@gmail.com

# I: Introduction

Predicting the thermodynamic properties and phase equilibria of multi-component fluids from pure fluid data is of immense importance in both academia and industry. This is typically achieved by use of an equation of state. For decades, cubic equations of state (EoS) such as Peng – Robinson (PR)[1] and SRK[2] have been the industry workhorses. In the latter part of the 20$^{th}$ century, perturbation theories[3,4] such as Weeks, Chandler, Andersen[3] and Barker-Hendersen[4] began to become more widely used. This class of perturbation theories assumes that fluid structure is dominated by repulsive forces[5] and attractions can be added as a perturbation in a high temperature expansion. Perturbation theories are grounded in statistical thermodynamics, giving a more sound theoretical footing as compared to their cubic counterparts.

Extension of perturbation theory to associating (hydrogen bonding) fluids was challenged by the strength and limited valence of the association interaction. Development of statistical mechanics for associating fluids was pioneered by the cluster expansions of Andersen[6] and Chandler and Pratt[7]. Application of perturbation theory to associating systems in an equation of state was not practical until the 1980's when Wertheim developed his multi-density statistical mechanics for associating fluids.[8,9] Wertheim's multi-density form of statistical mechanics allows for the development of relatively simple and accurate perturbation theories for associating fluids.[10–13] Perturbation theories based on Wertheim's multi-density statistical mechanics have come to be called "thermodynamic perturbation theory" or TPT for short. In TPT there is typically two energy scales: a short ranged highly direction association energy scale, and an orientationally averaged (non-association) energy scale. We loosely call this orientationally averaged contribution the reference energy scale. Typically, the reference energy scale will account for dispersion attractions, dipolar attractions not accounted for through the association term and higher order multi-pole contributions. When applied at first order in perturbation (TPT1), Chapman[14] developed a general form for the change in free energy due to association for fluids with any number of associating components. It is this form of thermodynamic perturbation theory which forms the basis of the statistical associating fluid theory (SAFT) family of engineering equations of state[15–18].

The true power of the SAFT equations of state for prediction and correlation of multi-component phase equilibria results from the accounting of multiple energy scales of intermolecular attraction. While traditional cubics such as PR and SRK only have a single energy scale of attraction, SAFT allows for separate accounting of hydrogen bonding and reference (non-hydrogen bonding) attraction degrees of freedom. This allows for much more accurate prediction of multi-component phase equilibria[19] in mixtures of polar and non-polar molecules. The challenge then becomes how to best choose the reference and hydrogen bonding energy scales to maximize the predictive capability of the model. In no system is the distribution of the total cohesive energy among hydrogen bonding and reference energy scales more important than mutual solubility predictions in water / hydrocarbon (HC) liquid-liquid equilibrium (LLE).

There have been numerous studies[20–22] where SAFT has been applied to study water / HC LLE. In an effort to deconvolute the reference and hydrogen bonding energy scales, Liang *et al.*[20] included water / HC LLE mutual solubility data, in addition to pure component water data, to obtain the pure component water parameters. It was shown that the resulting water parameter set accurately described the mutual solubilities of HC and water. However, the resulting parameter set loses the physical description of water with a chain length of $m = 2$ and segment diameter ~ 2.2 Angstrom. Similarly, there has recently been much attention[23–25] on using the solubility of water in the hydrocarbon phase to isolate the non-hydrogen bonding reference contribution to the water attractive energy. The logic being that when water is dilute in the hydrocarbon liquid phase it interacts solely with its reference attractive energy scale, hence the solubility of water in the HC phase can be used to isolate this energetic contribution.

It is known that water is highly polarized in the liquid state. A water monomer in the gas phase has a dipole moment of 1.855 D, while in the condensed phase this increases to 2.4 to 2.6 D (Debye) on average.[26] Through detailed first principles quantum mechanics calculations for small water clusters[27], it has been shown that water hydrogen bond energy, polarization and dipole moment[26,28,29] depend on the number of times the water molecule is hydrogen bonded, as well as the type of hydrogen bonded cluster. Hence, there is both hydrogen bond cooperativity as well cooperativity between hydrogen bonding and the non-hydrogen bonding (reference) energy scales. Logically, this means that the reference energy in TPT should depend on the degree of hydrogen bonding. Hence, it is not possible to isolate the reference energy scale.

In this paper we extend TPT to allow for the cooperativity of the reference energy scale with the hydrogen bonding energy scale. As our basis, we will use the PC-SAFT[17,30] form for the reference attractions and the standard TPT1 (SAFT)[15,17] treatment of the association attractions. We demonstrate that including this cooperativity in the theory allows for the prediction of water / HC LLE mutual solubilities. We accomplish this without the introduction of any additional parameters, and we do not use any binary data to tune the model.

## II: Theory

### A: Helmholtz free energy

In the SAFT EoS non-polar molecules are treated as chains of $m$ spheres of diameter $\sigma$ where the spheres attract with the reference energy $\varepsilon$. For hydrogen bonding molecules there are two additional parameters which describe the energy scale of hydrogen bonding $\varepsilon_{hb}$ as well as the association bond volume $\kappa_{hb}$. The Helmholtz free energy is given as the sum of a hard chain, reference attraction and hydrogen bonding contributions to the free energy

$$A = A_{hc} + A_{att} + A_{hb} \qquad (1)$$

For the hard chain contribution $A_{hc}$ we use the simplified pure component form as proposed by von Solms *et al* [31]. For the reference system attractions $A_{att}$ we employ the PC-SAFT form of Gross and Sadowski[30]. Finally, for the hydrogen bonding contribution we use the standard SAFT free energy contribution[15,17] with the pure component form of the hard sphere pair correlation function $g_{hs}$.[31] For a system with $N$ total molecules of $n_c$ species of mole fraction $x_k$, with each associating species having a set of association sites $\Gamma$, this contribution is given by[15]

$$\frac{A_{hb}}{Nk_bT} = \sum_{A \in \Gamma^{(k)}} \sum_{k=1}^{n_c} x_k \left( \ln X_A^{(k)} - \frac{X_A^{(k)}}{2} + \frac{1}{2} \right) \qquad (2)$$

The $X_A^{(k)}$ are the fraction of molecules of species $k$ NOT bonded at hydrogen bonding site $A$. These fractions are calculated self-consistently through the relations

$$\frac{1}{X_A^{(k)}} = 1 + g_{hs} \sum_{B \in \Gamma^{(j)}} \sum_{j=1}^{n_c} \kappa_{jk} \sigma_{jk}^3 \left( \exp\left( \frac{\varepsilon_{hb}^{(jk)}}{k_bT} \right) - 1 \right) \rho_j X_B^{(j)} \qquad (3)$$

### B: Temperature dependent hard sphere diameter

For classical simple fluids, it is the repulsive contribution to the intermolecular potential which dominates fluid structure.[5] These repulsive contributions to the free energy are modelled using results for a hard sphere fluid evaluated with a temperature dependent (or temperature and density) diameter $d_{jj}$. This approach accounts for the fact that real fluids have softer repulsions than that of hard spheres. The PC-SAFT attractive contribution is based on Barker-Hendersen perturbation theory and uses the following temperature dependent hard sphere diameter[30]

$$d_{jj} = \sigma_{jj}\left(1 - 0.12\exp\left(-3\frac{\varepsilon_{jj}}{k_b T}\right)\right) \quad (4)$$

From (4) we can see that increasing temperature decreases $d$ due to a softening of repulsions at high temperature. Replacing σ with $d$ in the hard sphere contributions will then have the effect of increasing density as compared to the case where the hard sphere diameter σ was used in these contributions. Again, this choice assumes that liquid structure is dominated by repulsive forces. However, for water, the fluid structure is also dependent on the geometries of the hydrogen bonds. In fact, due to cooperative effects, the average length of a hydrogen bond $R_{hb}$ decreases as the number of water molecules in the hydrogen bonded cluster increases.[32] The average number of hydrogen bonds per water molecules increases with a decrease in temperature which results in a shortening of the hydrogen bonds. Hence, for water, it would seem the use of a temperature dependent hard sphere diameter would have opposite the desired temperature dependence. In this work we propose the following relation for the temperature dependent diameter.

$$d_{jj} = \begin{cases} \sigma_{jj} & \text{for } j = water \\ \sigma_{jj}\left(1 - 0.12\exp\left(-3\frac{\varepsilon_{jj}}{k_b T}\right)\right) & \text{otherwise} \end{cases} \quad (5)$$

In (5) we have not attempted to add any new temperature dependence to the water diameter, we simply do not include the temperature dependent correction. As will be shown, Eq. (5) allows for substantial improvement in the theories ability to predict liquid water densities.

**C: Cooperativity among energy scales**

In all previous applications of SAFT, the reference system energy scale and the hydrogen bonding energy scale are considered to be independent, with the only coupling between the reference attraction and hydrogen bonding contributions to the free energy occurring through the temperature dependent hard sphere diameter $d_{jj}$. As has been noted, the degree of polarization of a water molecule depends on the hydrogen bonded state of the molecule.[27] It has been shown that this results in an increase in dipole moment[26,28,29] as a function of the degree of hydrogen bonding. On these bases we postulate that the reference system energy scale should depend on the degree of hydrogen bonding. To incorporate this effect in the theory, we consider each hydrogen bonded state of component $k$ to be a distinct sub-species $i$ with reference energy $\varepsilon_{kk}^{(ii)}$. In this development we consider $i$ to be the number of incident association bonds. Now applying van der Waals one-fluid mixing rules to these subspecies, we obtain the average reference energy of the hydrogen bonding component $k$ as

$$\varepsilon_{kk} = \sum_{i=0}^{n(\Gamma^{(k)})} \sum_{j=0}^{n(\Gamma^{(k)})} \varepsilon_{kk}^{(ij)} \chi_k^{(i)} \chi_k^{(j)} = \sum_{i=0}^{n(\Gamma^{(k)})} \sum_{j=0}^{n(\Gamma^{(k)})} \sqrt{\varepsilon_{kk}^{(ii)} \varepsilon_{kk}^{(jj)}} \chi_k^{(i)} \chi_k^{(j)} \tag{6}$$

Where $n(\Gamma^{(k)})$ is the number of association sites on species $k$, $\chi_k^{(i)}$ is the fraction of component $k$ which are hydrogen bonded $i$ times, and $\varepsilon_{kk}^{(ij)}$ is the cross attraction between a molecule of component $k$ bonded $i$ times and a molecule of component $k$ bonded $j$ times. For a given associating molecule, the fractions $\chi_k^{(i)}$ are calculated self-consistently through the hydrogen bonding fractions in Eq. (3). In the second step of Eq. (6) we assumed that the cross reference energy between subspecies $i$ and $j$ of component $k$ can be evaluated with a standard geometric mean mixing rule. To extend these ideas to mixtures, we factor out the monomer reference energy in Eq. (6) to obtain

$$\varepsilon_{kk} = \varepsilon_{kk}^{(00)} \theta_k^2 \tag{7}$$

where we have defined

$$\theta_k^2 = \sum_{i=0}^{n(\Gamma^{(k)})} \sum_{j=0}^{n(\Gamma^{(k)})} c_k^{(i)} c_k^{(j)} \chi_k^{(i)} \chi_k^{(j)} \quad ; \quad c_i^{(k)} = \sqrt{\frac{\varepsilon_{kk}^{(ii)}}{\varepsilon_{kk}^{(00)}}} \tag{8}$$

The cross interaction between components (not subspecies) $\varepsilon_{kl}$ is now calculated using the standard geometric mean mixing rule[30]

$$\varepsilon_{kl} = \sqrt{\varepsilon_{kk} \varepsilon_{ll}} (1 - k_{kl}) = \sqrt{\varepsilon_{kk}^{(00)} \varepsilon_{ll}^{(00)}} (1 - k_{kl}) \theta_k \theta_l \tag{9}$$

**D: Application to water**

We consider a 4 site water model with two hydrogen donor sites and two hydrogen bond acceptor sites. In TPT1 (which forms the basis of SAFT) each association bond is treated independently of all other association bonds on that molecule. This independence allows the fractions $\chi^{(i)}$ to be written as[33]

$$\chi^{(0)} = X_A^4 \quad ; \quad \chi^{(1)} = 4X_A^3(1 - X_A) \quad ; \quad \chi^{(2)} = 6X_A^2(1 - X_A)^2$$
$$\chi^{(3)} = 4X_A(1 - X_A)^3 \quad ; \quad \chi^{(4)} = (1 - X_A)^4 \tag{10}$$

All that remains is the determination of the coefficients $c_k^{(i)}$ in Eq. (8). As discussed above, the reference attraction between molecules will include dispersion as well as multi-pole attractions. It has been shown through detailed quantum calculations that the dipole moment and polarization of water increases with the degree of hydrogen bonding.[26,28,29,34,35] For instance, the dipole moment of a water monomer in the gas phase is 1.855 D, while calculations shows that the average dipole of individual water molecules is ~ 2.1 D in hydrogen bonded dimers and ~2.3 D in cyclic trimer clusters.[26] In this work we wish to employ insight gained from these detailed first principles calculations to estimate the coefficients in Eq. (8). To this end, we assume that the reference energies $\varepsilon_{kk}^{(ii)}$ have the same dependence on the degree of hydrogen bonding as the specific dipolar contribution to the intermolecular potential, which goes with the square of the dipole moment. On this basis we assume for $i = 0 - 2$ ($w$ = water)

$$c_{ww}^{(ii)} = \sqrt{\frac{\varepsilon_{ww}^{(ii)}}{\varepsilon_{ww}^{00}}} \approx \frac{\mu^{(i)}}{\mu^{(0)}} \tag{11}$$

where $\mu^{(i)}$ is the average dipole moment of a molecule bonded $i$ times. The dipole moment of water increases approximately linearly with the number of hydrogen bonds.[36] However, as noted by Glendening[37], the covalent character of the hydrogen bond means that orientations in hydrogen bonded clusters of water molecules will be dominated by the requirement that the orbitals of adjacent monomers by mutually orthogonal. That is, orientation will be optimized for hydrogen bonding, at the expense of a less than optimal configuration for multi-pole attractions. Hence, the benefit of larger polarization (in excess of the case $i = 2$) of more highly connected molecules may be offset by the fact that these more complex cluster geometries limit the availability of water molecules to interact with the dipoles of neighboring water molecules. Of course, this analysis does not account for increased polarization effects caused by the electric field of larger hydrogen bonded clusters.[34] See Mogensen and Kontogeorgis[38] for an analysis on the inclusion of long range polarization effects on dielectric properties.

In the current development, we have no way to self-consistently calculate liquid state structure (both spatial and orientational) based on the competition between hydrogen bonding and electrostatic attractions. Hence, for now, we assume that the coupling between reference energy and hydrogen bonding saturates at the $i = 2$ level meaning $\varepsilon_{ww}^{(22)} = \varepsilon_{ww}^{(33)} = \varepsilon_{ww}^{(44)}$. The constants for $i$

= 1 and 2 are then evaluated with Eq. (11) using dipole moment ratios consistent with first principles quantum calculations[26,28,29,34,35] on hydrogen bonded clusters as

$$c_w^{(1)} \approx \frac{\mu^{(1)}}{\mu^{(0)}} = 1.1 \quad ; \quad c_w^{(2)} \approx c_w^{(3)} \approx c_w^{(4)} \approx \frac{\mu^{(2)}}{\mu^{(0)}} = 1.2 \qquad (12)$$

Equation (12) completes the model for the coupling of reference and association energy scales in thermodynamic perturbation theory.

To summarize, we have used insights obtained from first principles calculations to propose a form the reference system energy as a function of the degree of hydrogen bonding in the system. We have neither introduced nor removed any parameters. The model still contains an empirical parameter to describe the reference system attractions; however, this parameter has now taken on a new meaning, as the reference contribution to the intermolecular potential of a water molecule which has no incident hydrogen bonds. Figure 1 plots the fraction of molecules bonded $k$ times, as well as the reference energy as a function of the overall fraction of bonded association sites $\chi = 1 - X_A$. As can be seen, the reference energy increases linearly with $\chi$ when hydrogen bonding is weak. At around $\chi \sim 0.5$ the rate of increase in the reference energy with $\chi$ begins to decrease; finally levelling out at a value $\varepsilon_{ww} = 1.44\varepsilon_{ww}^{(00)}$ when all water molecules are fully bonded.

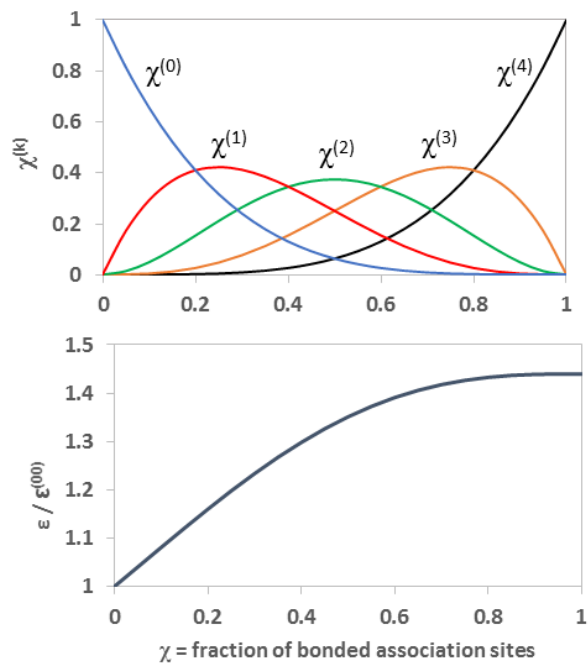

**Figure 1:** Fraction of molecules bonded k times Eq. (10) and reference energy Eq. (12) versus the fraction of bonded association sites for 4 site water model

## III: Results

In this section we apply the theory developed in II to pure water and water / hydrocarbon phase equilibria. To minimize the number of variables, we first treat water as a single sphere with a chain length $m = 1$. Second, we employ the results of Emborsky and Chapman(EC)[23] who obtained the reference system energy $\varepsilon_{ww}$ for water in PC-SAFT by regressing to water solubility in alkane data. As water is dilute in the hydrocarbon phase, their value of reference energy ($\varepsilon_{ww}$ = 209.84 $k_B$) will correspond to $\varepsilon_{ww}^{(00)}$. EC used a NIST equation of state for the water phase and PC-SAFT for the hydrocarbon phase; hence, their results should be very accurate. The remaining 3 parameters (hydrogen bond energy, hydrogen bond volume, hard sphere diameter) are obtained by data regression to saturated liquid densities ($\rho_L$) and vapor pressure ($P_{sat}$) data in the temperature range 273.15 – 580 K. We perform this analysis for two cases:

- Case I - No coupling $\varepsilon_{ww} = \varepsilon_{ww}^{(00)}$ and $d$ given by Eq. (4)
- Case II - Coupling between reference and association energy scales with $\varepsilon_{ww}$ given by Eqns. (6) and (12) and $d$ given by Eq. (5)

Case I gives the standard treatment while case II incorporates the modifications developed in this work.

Table I list the parameters obtained for each case as well as the average deviations obtained through the data regressions

**Table I:** Regression cases for water

| Case | $\sigma$ | $\kappa$ | $\varepsilon_{HB}/k_B$ | AAD% $\rho_L$ | AAD% $P_{sat}$ |
|---|---|---|---|---|---|
| Case I | 3.0661 | 0.04208 | 1899.3 | 4.1 | 2.3 |
| Case II | 3.0365 | 0.05646 | 1525.4 | 2.5 | 2.7 |

As can be seen, both cases give similar errors in the description vapor pressure; however, case II gives a substantial improvement in the overall description of the liquid densities (see Fig. 2). This improvement is a result of the use of the diameter $d$ given by Eq. (5). As discussed in II, the temperature dependent hard sphere diameter Eq. (4) assumes repulsions dominate the structure of the fluid; however, for water, hydrogen bonding also contributes. In water, as temperature is increased, there is a competition between softening of repulsions as described by Eq. (4), the breaking of hydrogen bonds, as well as the lengthening of the hydrogen bond distance[32] due to decreasing hydrogen bond cooperativity. This convolution of length scales (combined with the tetrahedral symmetry) makes it difficult for the equation of state to predict liquid water densities.

From Fig. 2 it can be seen that the use of the temperature dependent hard sphere diameter $d$ does more harm than good. The theory gives more accurate predictions of liquid density when the hard sphere diameter $\sigma$ is used. As can be seen, neither approach reproduces the density maximum of water. To predict this effect (on rigorous physical grounds) a theory would need to reproduce the tetrahedral structure of fully bonded water molecules. This is not possible in thermodynamic perturbation theory.

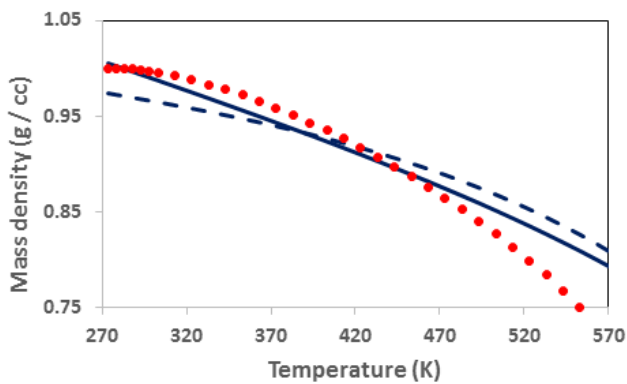

**Figure 2:** Model results (case I dashed curve and case II solid curve) and experimental data[39] (red circles) for the saturated liquid density of pure water.

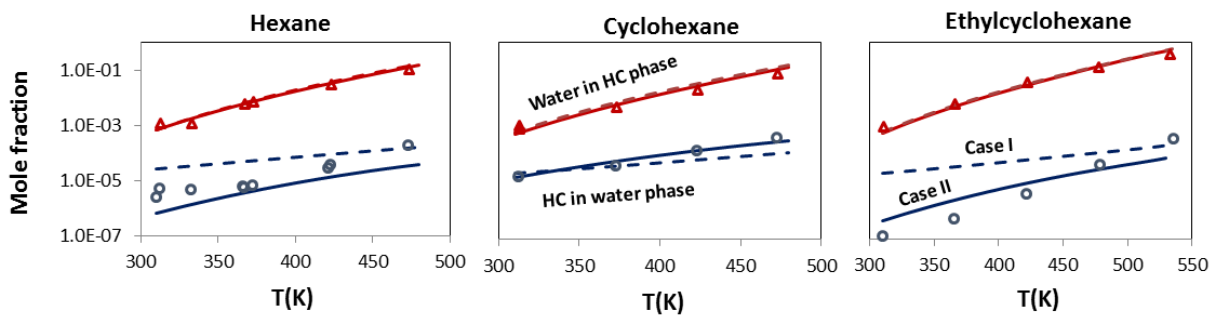

**Figure 3:** Mutual solubilities for water-hydrocarbon LLE. Dashed curves give case I model predictions, solid curves give case II model predictions and symbols are experimental data[40,41] (triangles water solubility in HC, circles HC solubility in water). Left: Hexane-water, Center: cyclohexane-water, Right: ethylcyclohexane-water

Figure 3 compares mutual solubility predictions ($k_{ij}$ = 0) and experimental data for water-hydrocarbon (HC) liquid-liquid equilibrium (LLE). As can be seen, both cases give a reasonable representation of the solubility of water in the HC phase. This should be expected, due to the fact that the energy $\varepsilon_{ww}^{(00)}$ was fit to solubility water solubility in HC data.[23] The test is then, how well does the theory take this information and predict the solubility of HC in the water phase? As can be seen, the new approach (case II) is much more successful. Case II gives substantially improved predictions for the solubility of HC in the water phase.

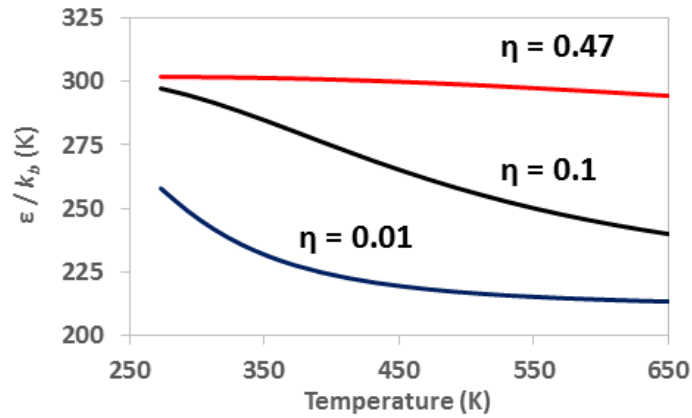

**Figure 4:** Reference energy of pure water versus temperature for several packing fractions $\eta = \pi \rho d^3 / 6$

This increase in predictive ability is the result of our coupling of reference and association energy scales in Eq. (6). This coupling can be further observed by the temperature and density dependence of the case II reference energy for pure water (Fig.4). As can be seen, the reference energy of water is a strong function of both density and temperature. Increasing temperature and decreasing density, results in a decrease in $\varepsilon_{ww}$ due to the breaking of hydrogen bonds. For mixtures, $\varepsilon_{ww}$ will also depend on the composition of the mixture. For instance, for the case of water-hexane LLE at 350 K (Fig. 3 left panel), the reference energy of water in the hexane rich phase where hydrogen bonding is limited is $\varepsilon_{ww}$ ~ 211 $k_B$, while it is substantially higher in the water phase where hydrogen bonding is extensive $\varepsilon_{ww}$ ~ 302 $k_B$. This also affects the cross attractions between water and hexane through the mixing rule Eq. (9), which predicts water-hexane attractions are stronger in the water rich phase than in the hexane rich phase. This seems counter intuitive due to the small solubility of hexane in water; however, small solubilities of hexane in the water rich phase are due to the fact that hexane cannot hydrogen bond and interferes with waters hydrogen bond network. This effect may help explain the failure[24,25] of classical mixing rules for water-alkane attractions.

It should be noted for case II that both the use Eq. (5) for $d$ and the inclusion of reference-hydrogen bond coupling complement each other. For instance, if one chooses only to use Eq. (5) for $d$, but does not include coupling of energy scales, refits water parameters as described in this section, very good results are obtained for liquid density predictions AAD = 1.35%; however, mutual solubilities with hydrocarbons are similar to those obtained for case I in Fig. 3. On the other hand, if one uses the temperature dependent diameter Eq. (4) for water and includes energy scale coupling, refits water parameters as described in this section, poor results are obtained for water liquid density predictions with an AAD = 5.7 %.

We conclude this section with a comparison to the experimentally measured fraction of free OH groups of Luck[42]. There has been much discussion in the literature on including Lucks data in the parameterization and application of SAFT.[43–45] Keeping with these we previous studies, we assume the fraction of free OH groups described by luck, is equivalent to the fraction of unbonded sites $X_A$. Figure 5 compares theory predictions to the measurements of Luck. Both theories over-predict the extent of hydrogen bonding (under-predict unbonded sites); however, the coupling of reference and hydrogen bonding energy scales results in a lower association energy for case II (table 1), which yields a better description of the data. While case II is likely within experimental error at low temperatures, the temperature dependence of the unbonded fractions is clearly too weak. As discussed in section I, it is known that water exhibits very strong hydrogen bond cooperativity. It is this authors opinion, that any accurate description of liquid water hydrogen bonding data over a wide temperature range, must include this effect. This is supported by the under-prediction of the temperature dependence in Fig. 5. Hydrogen bond cooperativity has been included in thermodynamic perturbation theory for two site associating fluids[46–48], however these specific approaches are not easily generalizable to hydrogen bonding fluids with $> 2$ association sites. For this, a new approach must be developed. This will be the subject of a forthcoming publication.

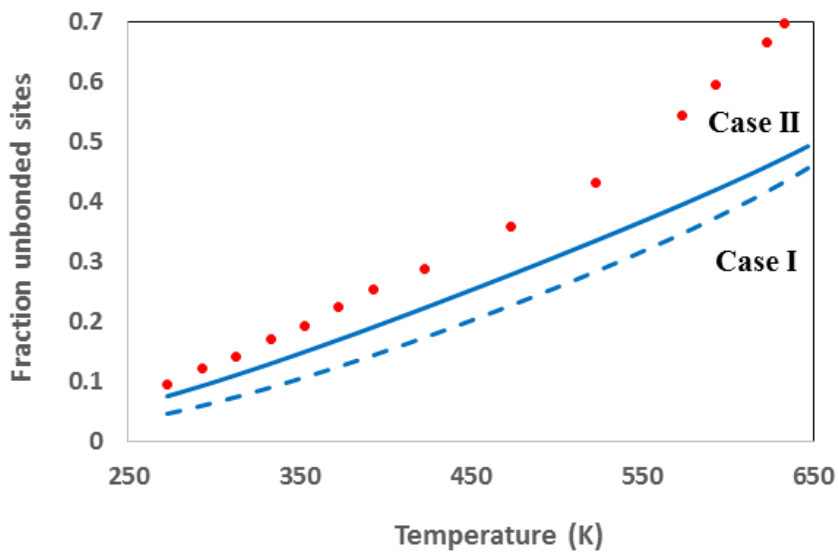

**Figure 5:** Comparison of fraction of unbonded association sites for case I (dashed curve) and case II (solid curve) to the spectroscopic data (circles) of Luck[42]

**IV: Discussion and Conclusions**

We have proposed a new methodology to couple the reference and hydrogen bonding energy scales in thermodynamic perturbation theory through the relation Eq. (6). To evaluate the unknown constants, we drew insight from the extensive body of literature on first principles calculations of the energetics of small water clusters. The resulting equation of state does not include any additional parameters as compared the original form. The new method was shown to give improved predictions of water liquid phase densities, as well as water-hydrocarbon mutual solubilities.

The results of this paper provide a simple modification of existing SAFT equations of state. However, the methodology is incomplete in a number of ways. First, the cooperativity between reference and association energy scales depends on the type of associated clusters (chains vs. rings for instance). In TPT1 we only allow for trees of hydrogen bonds. Also, in this work, there was a one way coupling between the reference and association energy scales. That is, while the degree of hydrogen bonding affects the reference energy scale, this dependence did not factor into the calculation of the degree of hydrogen bonding. To include this effect, one must return to the beginning, and include this dependence in the development of the multi-density[8] cluster expansion for associating fluids.

Finally, we have made no attempt to find the most accurate version of this model. An interesting next step would be to simultaneously fit all water pure component parameters (including $m$ and $\varepsilon_{ww}^{(00)}$) as well as the all constants $c_w^{(i)}$ ($i$ = 1-4) to experimental data. With this number of constants, liquid density and vapor pressure alone would likely prove insufficient. In addition to these standard data sources, one could also include heat capacity, heat of vaporization, vapor density, second virial coefficient, and possibly some binary phase equilibria data.